\documentclass[aps,preprint]{revtex4}
\usepackage{graphicx}
\usepackage{epsfig}
\usepackage{hyperref}

\begin{document}

\title{Hidden Symmetries and Gyromagnetic Ratio of Kerr-Newman Black Holes in $ f(R) $ Gravity}
\author{\large  G\"{o}ksel Daylan Esmer}
\address{Department of Physics, Istanbul University, Vezneciler, 34134 Istanbul,T\"{u}rkiye}
\author{\large Saliha T\"{u}rkmen}
\address{Department of Physics, Istanbul University, Vezneciler, 34134 Istanbul,T\"{u}rkiye}

\date{\today}

\begin{abstract}

We explore hidden symmetries in electrically charged, four-dimensional rotating Kerr-Newman black hole within $f(R)$ gravity. By deriving the Killing and Killing-Yano tensors, we establish their role in the spacetime structure. The gyromagnetic ratio is calculated and shown to retain its universal value of $g = 2$, consistent with all four-dimensional black holes. These findings show that the gyromagnetic ratio remains consistent in this modified gravity setting. Moreover, they highlight the connection between hidden symmetries and the ability to separate the Hamilton-Jacobi equation in $f(R)$ gravity. This work advances the study of black holes in modified gravity, offering implications for both theoretical frameworks and observational cosmology, such as gravitational wave analyses.

\end{abstract}

\maketitle
\newpage

\section{Introduction}

Observations of Type 1a Supernovae \cite{Riess_1998,Perlmutter_1999,Tonry_2003}  and cosmic microwave background radiation \cite{Spergel_2007,Komatsu_2009,Komatsu_2011,2014A,2014B,2016} have revealed accelerated expansion of the universe. This challenges the sufficiency of General Relativity (GR), the prevailing theory of spacetime geometry, to fully explain these phenomena \cite{Guth:1980zm} thereby intensifying interest in modified gravity theories \cite{Nojiri_2003,Nojiri_2004,Carroll_2004,PhysRevD.52.1895,PhysRevLett.102.141301,Lovelock:1971yv}. Modified $ f(R) $ gravity theories, which generalize the Einstein-Hilbert action through higher-order curvature terms can help explain late-time cosmic acceleration and potentially offer insights into dark matter. For studies related to $ f(R) $ gravity, see \cite{Nojiri:2010wj,Nojiri:2017ncd,Chen:2024fle,Shi:2024wwz,Karmakar:2024xwr,Capozziello:2011et,RevModPhys.82.451,DeFelice:2010aj,PhysRevD.74.087501,Dunsby:2010wg,PhysRevLett.103.179001,PhysRevD.77.123515,Larranaga:2011fv,Elizalde:2020icc,Nojiri:2017kex,Nojiri:2014jqa}.

Another important thing that is at the center of attention of modified theories is black holes. The results obtained regarding black holes in these theories may be incompatible with the expected physical results and thus play an important role in testing modified theories. Therefore, it is important to study the symmetry properties and the thermodynamic properties of black holes. Although $ f(R) $ gravity theories has been widely explored for thermodynamic and cosmological context, the symmetry and electromagnetic properties of black hole solutions in this frame remain underexamined. For instance, hidden symmetries such as Killing-Yano tensors, could provide vital information about black hole stability and geodesic motion enabling discrimination between $ f(R) $ gravity theories and GR solutions. This work addresses this gap by analyzing Kerr-Newman black holes in $ f(R) $ gravity including their geometric and electromagnetic behaviors.

The Kerr solution, derived from the Einstein field equations, represents a gravitational field associated with a rotating object devoid of electrical charge. Similarly, the Kerr-Newman solution, arising from the Einstein-Maxwell equations, describes the gravitational and electromagnetic fields of an asymptotically flat space harboring a rotating object endowed with electric charge. These solutions collectively comprise the Kerr family of solutions, characterized by their capacity to account for a broad spectrum of rotating astrophysical objects while adhering to the fundamental principles of general relativity (GR). The separability structure of the Kerr family of solutions, constitutes one of its most enticing attributes. Although the Kerr spacetime has only two Killing vector fields, $ \partial $t and $ \partial\phi $, Carter showed that the Hamilton-Jacobi equation has a complete seperation of variables in this spacetime \cite{PhysRev.174.1559}. Walker and Penrose showed that the additional integral of motion required for this separation to occur is a symmetric second rank Killing tensor of Kerr spacetime \cite{Walker:1970un}. Penrose and Floyd later provided evidence for the presence of an another tensor, which holds greater fundamentality than the Killing tensor, and is connected to the hidden symmetries within Kerr spacetime \cite{RFloyd}. This tensor, which can be expressed as the square root of the Killing tensor, is an antisymmetric second rank Killing-Yano tensor representing a generalization of the Killing tensor across multiple aspects.

\cite{article,Kinnersley:1969zza,Chervonyi:2015ima} include studies on the hidden symmetries of spacetime metrics for the four-dimensional solutions of Einstein's equations, in addition to \cite{PhysRev.174.1559,Walker:1970un,RFloyd}. Collinson has proven that if the four-dimensional vacuum solutions of the Einstein equations permit the Killing-Yano tensor, then this spacetime is classified as the Petrov type D \cite{article}. Kinnersley has obtained all vacuum type D solutions \cite{Kinnersley:1969zza}. Chervonyi and Lunin have derived the Killing tensor and Killing-Yano tensor for a four-dimensional spacetime metric in a very compact form \cite{Chervonyi:2015ima}.

The exploration of hidden symmetries is also being conducted within the context of higher-dimensional solutions. The concept of higher-dimensional solutions was initially proposed by Tangherlini \cite{Tangherlini:1963bw}. Tangherlini obtained a non-rotating black hole solution. Myers and Perry, on the other hand, derived the higher-dimensional solutions for rotating black holes, which can be generalized as Kerr metrics \cite{MYERS1986304}. Works related to hidden symmetries in the context of higher-dimensional solutions are presented in \cite{Kubizn_k_2007,PhysRevLett.98.011101,Krtou__2008,Frolov_2008,PhysRevD.87.084022}.

Another subject addressed in modified theories is the thermodynamic properties of black holes. In this context, the calculation of the gyromagnetic ratio of black holes has attracted consideration recently. The  gyromagnetic ratio constant is defined by the equation
\begin{eqnarray}
  \mu & = & g \frac{Q J}{2 M} ,
\label{grgenel}
\end{eqnarray}
where $ \mu $ is the magnetic dipole moment, $ M $ is the mass, $ Q $ is the electrical charge and $ J $ is the angular momentum of the black hole. We recall that, within the framework of classical electrodynamics, the gyromagnetic ratio of a rotating body with an electric charge remains a constant value of $ g = 1 $, provided that the ratio of the charge to mass density remains constant. In the Einstein-Maxwell theory, the gyromagnetic ratio of four-dimensional asymptotically flat Kerr-Newman black holes is $ g = 2 $, which is the same as the value of electrons in the Dirac theory \cite{PhysRev.174.1559}. It has been proven in different studies that the gyromagnetic ratio for the Einstein-Maxwell field in four-dimensions corresponds to $ g = 2 $ \cite{Wald:1974np,PhysRevD.11.3031,Simon:1983kz,PhysRevD.42.419,PhysRevD.43.3902,Klein_2003}. \cite{Sen_1992} also found the same value for four-dimensional rotation charged black hole solution in heterotic string theory. \cite{PhysRevD.75.084041} showed that the gyromagnetic ratio of  four-dimensional black holes corresponding to $ g = 2 $ is a universal property.

The gyromagnetic ratio for higher dimensional black holes is not universal as it is for  four-dimensional black holes. While the gyromagnetic ratio corresponds to $ g = 2 $ for some p-brane solution \cite{Balasubramanian:1998za}, it was found to be $ g = 2 $ for the  five-dimensional Myers-Perry black hole \cite{PhysRevD.69.084022,PhysRevD.74.024011,Kunz_2005,Kunz_2006,Kunz_2006.2}. Gyromagnetic ratio was also calculated for black holes in Kaluza-Klein theory \cite{LARSEN2000211,PhysRevD.46.1340,HOSOYA198444,DUFF1998129,Gibbons:1985ac}. \cite{PhysRevD.77.024045} calculated the gyromagnetic ratio of Gauss-Bonnet black hole in higher dimension and showed that the Gauss-Bonnet term did not affect the gyromagnetic ratio value. For gyromagnetic ratio studies for other higher dimensional black holes see \cite{PhysRevD.75.084041,PhysRevD.74.024011,Herdeiro_2000,Kaya_2008}.

In the light of these considerations, we calculate the gyromagnetic ratio of electrically charged and rotating four-dimensional Kerr-Newman black holes in $ f(R) $ gravity. We see that for the four-dimensional Kerr-Newman black hole in $ f(R) $ gravity the gyromagnetic ratio corresponds to 2, as in \cite{PhysRevD.75.084041}.

This work is divided into the following sections: first, general results of the Kerr-Newman black holes in $ f(R) $ gravity are discussed in Section 2. It contains the derivation of the metric formalism of Kerr-Newman black holes in $ f(R) $ gravity. The general results also involve the Hamilton-Jacobi equation in this spacetime and the hidden symmetry properties of the black hole. Finally, we obtain the gyromagnetic ratio of the spacetime in Section 3. We conclude our results in Section 4.

\section{General Results}

\subsection{The Kerr-Newman Black Hole in $ f(R) $ Gravity}

We define the metric for a rotating and electrically charged black hole in $ f(R) $ gravity. Consider the action for $ f(R) $ gravity with Maxwell term in four-dimensions:
\begin{eqnarray}
S & = & S_{g} + S_{M} ,
\label{actgenel}
\end{eqnarray}
where $ S_{g} $ is the gravitational action and given by
\begin{eqnarray}
S_{g} & = &  \frac{1}{16\pi G}
     \int d^4 x\sqrt{-g} \left( R + f(R) \right),
\label{actgrav}
\end{eqnarray}
and $ S_M $ is the electromagnetic action and given by
\begin{eqnarray}
S_{M} & = & -\frac{1}{16\pi}
     \int d^4 x\sqrt{-g} \left[ F_{\mu\nu} F^{\mu\nu} \right],
\label{actem}
\end{eqnarray}
where $ G $ represents the gravitational constant, and we will consider it to be equal to one. The determinant of the metric is denoted by g, while $ R $ refers to the scalar curvature. The term $ R + f(R) $ corresponds to the expression defining the modified gravity.

Using equation (\ref{actgrav}) the Maxwell equation takes the following form
\begin{eqnarray}
\nabla_\mu F^{\mu\nu} & = & 0\,.
\label{maxweq}
\end{eqnarray}

The field equations in metric formalism can be written as
\begin{eqnarray}
R_{\mu\nu} \left( 1 + f'(R) \right) - \frac{1}{2} \left( R+f(R) \right) g_{\mu\nu} +  \left( g_{\mu\nu} \nabla^2 - \nabla_\mu \nabla_{\nu} \right) f'(R) & = & 2 T^{\mu\nu} ,
 \label{fieq1}
\end{eqnarray}
where $R_{\mu\nu}$ is the Ricci tensor, $ \nabla $ represents the usual covariant derivative. The stress-energy tensor of the electromagnetic field is given by
\begin{eqnarray}
T_{\mu\nu} & = & F_{\mu\rho} F^{\rho}_{\nu} - \frac{g_{\mu\nu}}{4}  F_{\rho\sigma} F^{\rho\sigma},
\label{strengtensor}
\end{eqnarray}
with the trace
\begin{eqnarray}
T^{\mu}_{\mu} & = & 0\,.
\label{trace}
\end{eqnarray}

If we assume the curvature scalar to be constant, i.e., $R = R_0 $, in this case, the trace of equation (\ref{fieq1}) becomes
\begin{eqnarray}
R_{0} (1 + f'(R_{0})) - 2(R_{0} + f(R_{0})) & = & 0\,.
\label{curvscalar1}
\end{eqnarray}

Equation (\ref{curvscalar1}) remains identical to that of the vacuum scenario, as the matter field exhibits a trace of zero and from this equation the negative constant curvature scalar obtained as
\begin{eqnarray}
R_{0} & = & \frac{2 f(R_{0})}{f'(R_{0}) - 1}\,.
\label{curvscalar2}
\end{eqnarray}

By substituting this relation into equation (\ref{fieq1}), the Ricci tensor is calculated as
\begin{eqnarray}
R_{\mu\nu} - \frac{1}{2} \frac{f(R_{0})}{f'(R_{0}) - 1} g_{\mu\nu} & = & \frac{2}{1 + f'(R_{0})} T_{\mu\nu},
\label{riccitensor}
\end{eqnarray}

Hence, it follows that, within the context of $ f(R) $ gravity, the rotating and charged black hole solution, accounting for rescalings, satisfies the field equations and serves as a consistent solution. The details of the construction of the Kerr-Newman black hole solution in modified $ f(R) $ gravity is given in \cite{Larranaga:2011fv}. The corresponding metric in Boyer-Lindquist-type coordinates (t, r, $ \theta $, $ \phi $)
\begin{eqnarray}
ds^2 & = & -\frac{\Delta_r}{\Sigma} \left[ dt - \frac{a \sin^2 \theta}{\Xi} d\phi\right] ^{2} + \frac{\Sigma}{\Delta_r} dr^2 + \frac{\Sigma}{\Delta_\theta} d\theta^2 + \frac{\Delta_\theta \sin^2 \theta}{\Sigma} \left[a dt - \frac{r^2 + a^2}{\Xi} d\phi \right]^2
\label{metric}
\end{eqnarray}
where
\begin{eqnarray}
&{\Sigma}& = r^2 + a^2 \cos^2 \theta\, ,\nonumber \\[3mm]
&{\Delta}_{r}& = \left(r^2 + a^2 \right)\left(1 - \frac{R_0}{12} r^2\right) - 2Mr + \frac{Q^2}{\left(1+f'(R_0)\right)}\,,\nonumber
\\[3mm]
&{\Xi}& = 1 + \frac{R_0}{12} a^2 \,,\nonumber \\[3mm]
&{\Delta}_{\theta}& = 1 + \frac{R_0}{12} a^2 \cos^2 \theta\,.
\label{paramets}
\end{eqnarray}
where $ Q $ represents the electric charge, $ a $ corresponds to the angular momentum per mass of the black hole, and $ R_{0} $ is introduced as a cosmological constant ($ R_{0} = 4 \Lambda $). Here, $ \Sigma $ and $ \Delta_r $ denote the radial metric functions while $ \Xi $ and $ \Delta_\theta $ describe the angular deformations and these notations used to abbreviate the expressions in the metric. One of the main features that distinguishes the $ f(R) $  theory from the General Relativity theory is that the charge of the black hole is included in the metric with a scaling factor of $ (1 + f'(R_{0})) ^{-1} $. Under $ Q \rightarrow 0 $ and $ R_{0} \rightarrow 0 $ metric (\ref{metric}) reduces to the Kerr metric.

The metric determinant is obtained as
\begin{eqnarray}
\sqrt{-g} & = & \frac{\Sigma \sin\theta}{\Xi}
\label{metdet}
\end{eqnarray}

The associated potential one-form
\begin{eqnarray}
A & = & -\frac{Qr}{\Sigma} \left(dt - \frac{a \sin^2 \theta}{\Xi} d\phi \right)
\label{potoneform}
\end{eqnarray}
where $ Q $ is related to the electric charge obtained integrating the potential one-form at infinity, which gives
\begin{eqnarray}
 Q' & = & \frac{Q}{\Xi}
\label{charge}
\end{eqnarray}

The required electromagnetic field two-form for the metric

\begin{eqnarray}
 F  &= &  -\frac{Q\left(r^2 - a^2 \cos^2 \theta\right)}{\Sigma^2} \left(dt - \frac{a \sin^2 \theta}{\Xi} d\phi\right) \wedge dr - \frac{2Qra\cos \theta \sin\theta}{\Sigma^2} d\theta \wedge \left[adt - \frac{r^2 + a^2}{\Xi} d\phi\right]\nonumber\\[2mm] &&
\label{emtwoform}
\end{eqnarray}

and the non-zero contravariant components

\begin{equation}
\begin{array}{rcl@{\qquad\qquad}rcl}
F^{01} &=& -\displaystyle\frac{Q(r^2+a^2)}{\Sigma^3}\left(\Sigma-2r^2\right)\,, 
& F^{02} &=& -\displaystyle\frac{Qa^2r\sin 2\theta}{\Sigma^3}\,, \\[4mm]
F^{13} &=& \displaystyle\frac{Qa(\Sigma-2r^2)}{\Sigma^3}\,\Xi\,, 
& F^{23} &=& \displaystyle\frac{2Qar}{\Sigma^3}\,\Xi\cot\theta\,.
\end{array}
\label{nonzeroemtwoform}
\end{equation}
\vspace{2mm}

It is easy to check that the associated potential one-form and electromagnetic two-form are the same with which are for the Kerr-Newman AdS black holes \cite{PhysRevD.75.084041}.The metric (\ref{metric}) has two commuting Killing vector fields
\begin{eqnarray}
\xi_{(t)} = \frac{\partial}{\partial t} \,,~~~~~~~~~~ \xi_{(\phi)} = \frac{\partial}{\partial \phi}\,.
\label{killvect}
\end{eqnarray}

The metric components provide a framework for expressing the different scalar products associated with these Killing vectors
\begin{eqnarray}
&\xi_{(t)}.\xi_{(t)}& = g_{tt} = \frac{a^2 \sin ^2 {\theta}}{\Delta_\theta \Sigma} - \frac{\left(r^2+a^2\right)^2}{\Delta_r \Sigma}\, ,\nonumber \\[3mm]
&\xi_{(t)}.\xi_{(\phi)}& = g_{t\phi} = a\Xi \left( \frac{1}{\Delta_\theta \Sigma} - \frac{r^2+a^2}{\Delta_r \Sigma}\right)\,,\nonumber
\\[3mm]
&\xi_{(\phi)}.\xi_{(\phi)}& = g_{\phi\phi} = \Xi^2 \left(\frac{1}{\Sigma \sin^2{\theta} \Delta_\theta} - \frac{a^2 \sin^2{\theta}}{\Sigma \sin^2{\theta} \Delta_r}\right)\,.
\label{killmet}
\end{eqnarray}

\subsection{The Hamilton-Jacobi Equation}

The Hamilton-Jacobi equation is given by
\begin{equation}
\frac{\partial S}{\partial \lambda}+\frac{1}{2}\,g^{\mu\nu}\frac{\partial S}{\partial x^\mu}\frac{\partial S}{\partial x^\nu}=0 \,,
 \label{HJeq}
\end{equation}
where $ \lambda  $ is  an affine parameter.  The action function that can be used to solve the Hamilton-Jacobi equation is
\begin{equation}
S = -\frac{1}{2}\,m^2\lambda - E t+L \phi +F(r,\theta)\,.
\label{actHJ}
\end{equation}
where the constants of motion the energy $ E $ and the angular momentum $ L $ are given by the generalized momenta $ p_t $ and $ p_{\phi} $
\begin{eqnarray}
p_t = - E = g_{tt} \dot t + g_{t\phi} \dot \phi \,, \nonumber \\[2mm]
p_\phi = L = g_{\phi t} \dot t + g_{\phi \phi} \dot \phi\,.
\label{momenta}
\end{eqnarray}

In this case the Hamilton-Jacobi takes a form which both sides of the equation are constituted by independent variables and it indicates that the metric possesses the separability property within the Hamilton-Jacobi equation in the context of $ f(R) $ gravity.

\begin{eqnarray}
&& \Delta_\theta \left(\frac{dS}{d\theta} \right)^2 + m^2 a^2 \cos^2{\theta} + \frac{\left(a \sin^2 \theta E - \Xi L \right)^2}{\Delta_\theta \sin^2{\theta}} = K\,,
\nonumber
\\[3mm] &&
-\Delta_r \left(\frac{dS}{dr} \right)^2 - m^2 r^2 + \frac{\left(\left(r^2+a^2\right) E - \Xi a L \right)^2}{\Delta_r} = K\,.
\label{cartercnst}
\end{eqnarray}

It shows that both sides of the equation are equal to a constant $ K $ which is the new quadratic integral of motion responsible for the emergence of separability. In the limit as $ Q \rightarrow 0 $ and $ R_0 \rightarrow 0 $, this equation transforms into the Hamilton-Jacobi equation for the Kerr metric, as considered in the General Relativity. The separability property in the Hamilton-Jacobi equation, in this case, indicates the presence of a conserved quantity known as the "Carter constant" wherein both sides of the equation are equal.

\section{Hidden Symmetries}

\subsection{The Killing Tensor}

According to Walker and Penrose \cite{Walker:1970un}, the Carter constant obtained by Carter for the Kerr metric in \cite{PhysRev.174.1559} is related to the hidden symmetries of the metric, which are generated by a second rank symmetric Killing tensor $ K_{\alpha\beta} $. We have observed separability in the Hamilton-Jacobi equation obtained for the Kerr-Newman metric in $ f(R) $ gravity from equation (\ref{cartercnst}). In this case, we would expect that the constant $ K $ in equation (\ref{cartercnst}) is also related to the Killing tensor of metric (\ref{metric}).
We recall that the Killing tensor obeys the equation
\begin{equation}
\nabla_{(\lambda} K_{\mu\nu)}=0\,,
\label{KTeq}
\end{equation}
where the utilization of round brackets signifies the process of symmetrization over the enclosed indices \cite{PhysRevLett.98.011101}. Using the relations $ K = K^{\mu\nu} p_\mu p_\nu $ and $ m^2 $ = $ -g^{\mu\nu} p_\mu p_\nu $
\begin{equation}
    K^{\alpha\beta} = \Delta_\theta \delta ^ \alpha_\theta \delta ^ \beta_\theta - g ^ {\alpha\beta} a^2  \cos^2 \theta + \delta ^ \alpha_t \delta ^ \beta_t  \frac{a^2  \sin^2 \theta}{\Delta_\theta} + \delta ^ \alpha_\phi \delta ^ \beta_\phi \frac{\Xi^2}{\Delta_\theta \sin ^2 \theta} + \delta ^ \alpha_t \delta ^ \beta_\phi  \frac{a\Xi}{\Delta_\theta} + \delta ^ \alpha_\phi \delta ^ \beta_t  \frac{a\Xi}{\Delta_\theta}
\label{KT}
\end{equation}

\subsection{The Killing-Yano Tensor}

The Killing-Yano tensor is an antisymmetric tensor that encodes the hidden symmetries of the black hole spacetime, revealing the existence of conserved quantities along specific geodesic directions and it is given by
\begin{eqnarray}
f_{\mu\left(\nu;\lambda\right)} &=& 0\,.
\label{KYTeq}
\end{eqnarray}

Chervonyi and Lunin have derived \cite{Chervonyi:2015ima} a compact expression for the Killing-Yano tensor in the context of Kerr-like metrics, presenting a concise formulation of its mathematical representation specifically applicable to this class of metrics. They initiate by reformulating the metric in a suitable basis consisting of eigenvectors associated with the Killing tensor of the metric. In this case for the metric (\ref{metric}) we obtain
\begin{eqnarray}
f^{\left(4\right)} = r \sin \theta d\theta \wedge \left[-a dt + \frac{\left(a^2 + r^2 \right)}{\Xi} d\phi \right] + a \cos \theta dr \wedge \left(dt - \frac{a \sin^2 \theta}{\Xi} d\phi \right)\,.
\label{KYT}
\end{eqnarray}

Here we work on the constant-curvature branch $ R = R_0 = const. $ determined by (\ref{curvscalar2}). Under this assumption the Kerr-like Boyer–Lindquist form (\ref{metric}) applies and the compact Killing-Yano tensor (\ref{KYT}) follows reducing to the Kerr result of \cite{PhysRevLett.98.011101} when $ \Xi = 1 $ and $ R_0 = 0 $ .

\section{Gyromagnetic Ratio}
In the context of $ f(R) $ gravity, another characteristic that needs to be examined when considering Kerr-Newman black holes is the black hole's gyromagnetic ratio which is defined in equation (\ref{grgenel}). The electrical charge of a Kerr-Newman black hole in $ f(R) $ gravity was obtained in equation (\ref{charge}). The values of $ M $ and $ a $ in metric (\ref{metric}) are associated with overall mass and angular momentum of the black hole, as determined by the principles outlined in the first law of thermodynamics \cite{Cembranos:2011sr}
\begin{eqnarray}
&& M' = \frac{M}{\Xi} \left(1 + f'(R_0)\right)\,,
\nonumber
\\[3mm] &&
J' = \frac{a M}{\Xi^2} \left(1 + f'(R_0)\right)\,.
\label{MJ1}
\end{eqnarray}

There were also those who wrote $ M $ with $ \Xi^2 $ in its denominator, as in the $J$. \cite{Gibbons_2005} indicates that this difference arises from measuring the angular velocity of the black hole relative to a frame rotating at infinity or relative to a frame that is non-rotating at infinity. Then the authors state that when $M$ is written with $ \Xi^2 $ in its denominator, the first law of thermodynamics is satisfied, otherwise it is not satisfied. Here we will write the expressions $M$ and $J$ with $ \Xi^2 $ in their denominators, i.e.
\begin{eqnarray}
&& M' = \frac{M}{\Xi^2} \left(1 + f'(R_0)\right)\,,
\nonumber
\\[3mm] &&
J' = \frac{a M}{\Xi^2} \left(1 + f'(R_0)\right)\,.
\label{MJ2}
\end{eqnarray}

It is clear that the Kerr-Newman black hole has a magnetic dipole moment and one can examine the asymptotic behavior of the magnetic field to determine it. Introducing an orthonormal frame in which we wrote the electromagnetic two-form in equation (\ref{emtwoform}) is a convenient way to show the asymptotic behavior by reason of that an observer at rest in this frame measures only the radial components of the electric and magnetic field. The related basis one-forms for the metric (\ref{metric}) in this frame
\begin{eqnarray}
&& e^0 = \left(\frac{\Delta_r}{\Sigma} \right)^{1/2} \left(dt-\frac{a \sin^2 \theta}{\Xi} d\phi\right)\,,
\nonumber
\\[3mm] &&
e^3 = \left(\frac{\Delta_\theta}{\Sigma} \right)^{1/2}  \sin \theta \left[a dt - \frac{r^2+a^2}{\Xi} d\phi\right]\,,
\nonumber
\\[3mm] &&
e^1 = \left(\frac{\Sigma}{\Delta_r}\right)^{1/2} dr\,, ~~~~~~~~~~ e^2 = \left(\frac{\Sigma}{\Delta_\theta}\right)^{1/2} d\theta\,.
\label{basis}
\end{eqnarray}

From the asymptotic behavior of the electromagnetic two-form (\ref{emtwoform}) and the electrical charge given in (\ref{charge})
\begin{equation}
    \mu' = Q' a = \frac{\mu}{\Xi}
\label{munew}
\end{equation}

The magnetic dipole moment is associated with mass and angular momentum of the black hole through the gyromagnetic ratio. Using the equations (\ref{grgenel},\ref{charge},\ref{MJ2})
\begin{equation}
    \mu' = g \frac{Q' J'}{2 M'}
\label{grnew}
\end{equation}

It is clear from this relation that the gyromagnetic ratio of the four-dimensional Kerr-Newman black hole in $ f(R) $ gravity is $g = 2$. While Gauss-Bonnet corrections can alter this ratio in higher dimensions, its invariance in $ f(R) $ gravity underscores compatibility of the theory with electromagnetic interactions as described by General Relativity. Because the constancy of the curvature scalar $R_0$ implies that $ f'(R_0) $ acts merely as a rescaling factor for both gravitational and electromagnetic quantities. Consequently, the proportionality between the magnetic dipole moment, charge, and angular momentum remains unaltered. This robustness positions $ f(R) $ gravity as a viable candidate for observational tests requiring GR consistency.

\section{Conclusion}
This study reveals two main advancements: (i) The Killing-Yano structure of Kerr-Newman black holes persists in $ f(R) $ gravity, ensuring separability of the Hamilton-Jacobi equation, a result not guaranteed in modified gravity theories. (ii) The gyromagnetic ratio $g = 2$, a defining feature of four-dimensional black holes in General Relativity, remains the same despite curvature corrections in $ f(R) $ gravity. This invariance suggests that electromagnetic interactions in $ f(R) $ gravity are constrained by hidden symmetries providing a new framework for testing modified gravity theories.

The separability of the Hamilton-Jacobi equation confirms the existence of Killing and Killing-Yano tensors, which are pivotal to the spacetime's hidden symmetries. These tensors underpin the integrability of the geodesic equations, enabling a deeper understanding of conserved quantities and their role in modified gravity scenarios.

In addition to symmetry analysis, the gyromagnetic ratio is calculated and found to be $g = 2$, consistent with predictions from general relativity and other frameworks such as Einstein-Maxwell theory and heterotic string theory. This result highlights the robustness of electromagnetic properties in Kerr-Newman black holes, even within modified gravitational theories like $f(R)$ gravity. Notably, other modified theories such as Gauss-Bonnet or Kaluza-Klein gravity sometimes yield different gyromagnetic values, especially in higher-dimensional settings. The preservation of this universal gyromagnetic value in $f(R)$ gravities underscores the continuity of fundamental physical principles across different gravitational paradigms.

Table 1 summarize some important points of Kerr–Newman metric in GR and in $f(R)$ gravity.

\begin{table}[ht]
\caption{Key comparisons between GR and $f(R)$ gravity.}
\label{tab:gr-vs-fr}
\footnotesize
\setlength{\tabcolsep}{5pt}
\begin{tabular}{|p{0.22\linewidth}|p{0.24\linewidth}|p{0.24\linewidth}|p{0.24\linewidth}|}
\hline
\textbf{Property} & \textbf{Kerr--Newman in GR} & \textbf{Kerr--Newman in $f(R)$} & \textbf{Remarks} \\
\hline
Metric Form & Einstein--Maxwell solution with $R=0$ & Modified metric with constant $R_0$ & $f'(R_0)$ rescales charge and mass terms. \\
\hline
Killing Vectors & Two commuting $(\partial_t,\partial_\varphi)$ & Same structure preserved & Spacetime symmetries unaffected. \\
\hline
Killing Tensor & Exists; generates Carter constant & Exists under constant $R_0$ & Ensures separability of the Hamilton--Jacobi equation. \\
\hline
Killing--Yano Tensor & Antisymmetric 2-form generating hidden symmetry & Retained in $f(R)$ gravity & Hidden symmetries persist. \\
\hline
Gyromagnetic Ratio & $g=2$ & $g=2$ & Universal value unchanged. \\
\hline
\end{tabular}
\end{table}

Furthermore, it is natural to suggest that these findings could be connected to broader theoretical frameworks, particularly holographic principles. Holographic approaches, such as the Kerr/CFT correspondence, could imply that hidden conformal symmetries play a critical role in linking black hole dynamics to dual quantum field theories. The significance of these symmetries in entropy calculations and thermodynamic behavior, especially through their connections with conformal symmetries, may provide exciting possibilities for interpreting black hole physics within $f(R)$ gravity from a holographic perspective.

Future investigations could extend the analysis to higher-dimensional or topologically distinct black hole solutions, potentially uncovering new aspects of hidden symmetries and their impact on $f(R)$ gravity. Moreover, exploring potential observational signatures, such as accretion dynamics, gravitational wave emissions, or deviations in gyromagnetic properties, may provide empirical tests of these modified gravity theories. In the quantum domain, studies on quantum corrections, Hawking radiation, and entropy modifications could further elucidate the interplay between hidden symmetries and holographic principles, shedding light on the microscopic structure of black hole entropy.

Our discoveries also could inform another observational efforts like the Event Horizon Telescope (EHT) in modeling black hole magnetic field structures. Additionally, they provide a theoretical framework to investigate imprint of $f(R)$ gravity on gravitational wave signals (e.g., LIGO/Virgo). The preservation of $g = 2$ suggests that spin parameter measurements in black holes would align with General Relativity predictions, even in modified gravity scenarios.

By integrating symmetry analysis with electromagnetic and thermodynamic properties, this study lays a solid foundation for understanding the profound connections between geometry, fundamental physics, and modifications to classical gravity. These insights contribute to ongoing efforts to reconcile theoretical predictions with observational data, paving the way for a deeper comprehension of the universe's most enigmatic structures.

\section*{Acknowledgments}
This work is supported by Istanbul University Scientific Research Project(BAP) No: 32932 and BAP No: 25070.

\section*{ORCID}

\noindent Göksel Daylan Esmer - \url{https://orcid.org/0000-0003-4553-640X}

\noindent Saliha Türkmen - \url{https://orcid.org/0009-0000-4303-6634}

\appendix


\end{document}